\documentclass[10pt]{article}
\usepackage{graphicx,float,subeqnarray,colordvi}
\usepackage{amsmath,amsfonts,amssymb,latexsym}

\def\R{\mathbb{R}}
\def\C{\mathbb{C}}
\def\tL{\widetilde{L}}
\def\tM{\widetilde{M}}
\def\rT{{\rm T}}
\def\ts{\widetilde{\sigma}}
\def\lm{\lambda_{\max}}
\def\imi{\textbf{\hskip1pt i\hskip1pt}}

\newtheorem{theorem}{Theorem}

\newtheorem{lemma}[theorem]{Lemma}

\date{October 7, 2010}

\begin{document}
\title{Stability of central finite difference schemes\\ for the Heston PDE}
\author{K.~J.~in 't Hout\thanks{Department of Mathematics and Computer Science,
University of Antwerp, Middelheimlaan 1, B-2020 Antwerp, Belgium
(\mbox{e-mail}: \texttt{karel.inthout@ua.ac.be}).}
~and K.~Volders\thanks{Department of Mathematics and Computer Science,
University of Antwerp, Middelheimlaan 1, B-2020 Antwerp, Belgium
(\mbox{e-mail}: \texttt{kim.volders@ua.ac.be}).}}
\maketitle

\begin{abstract}
This paper deals with stability in the numerical solution of 
the prominent Heston partial differential equation from 
mathematical finance.
We study the well-known central second-order finite difference 
discretization, which leads to large semi-discrete systems with
non-normal matrices $A$.
By employing the logarithmic spectral norm we prove practical, 
rigorous stability bounds.
Our theoretical stability results are illustrated by ample 
numerical experiments.
\end{abstract}
\vspace{0.2cm}

stability, contractivity, 
logarithmic norm.
\vspace{3mm}

\vfill\eject

\setcounter{equation}{0}
\setcounter{theorem}{0}

\section{Introduction}
This paper deals with stability in the numerical solution of the Heston 
partial differential equation (PDE), 
\begin{equation}\label{Heston}
\frac{\partial u}{\partial t}
=\tfrac{1}{2}s^2v\frac{\partial^2 u}{\partial s^2}
+\rho\sigma sv\frac{\partial^2 u}{\partial s\partial v}
+\tfrac{1}{2}\sigma^2v\frac{\partial^2 u}{\partial v^2}
+rs\frac{\partial u}{\partial s}
+\kappa(\eta-v)\frac{\partial u}{\partial v}-ru
\end{equation}
for $s>L$,~$v>0$ and $0<t\leq T$. 
The Heston PDE constitutes one of the prominent equations of mathematical 
finance, cf.~e.g.~\cite{H93,L01,S04,TR00}.
It generalizes the celebrated one-dimensional Black--Scholes PDE 
where the volatility is modelled by a stochastic process rather than 
being constant.
Clearly, (\ref{Heston}) can be viewed as a time-dependent 
advection-diffusion-reaction equation on an unbounded two-dimensional 
spatial domain. 
The exact solution value $u(s,v,t)$ represents the fair 
price of a European-style option if at time $T-t$ the underlying asset price 
and its variance equal $s$ and $v$, respectively, where $T>0$ is the given 
maturity time of the option.
The quantity $L\ge 0$ is a lower barrier, $\kappa>0$ is the mean-reversion 
rate, $\eta>0$ is the long-term mean, $\sigma>0$ is the volatility-of-variance, 
$\rho\in[-1,1]$ is the correlation between the two underlying Brownian motions, 
and $r>0$ is the interest rate.
These quantities are all given and arbitrary.
We remark that in practice the correlation $\rho$ is usually nonzero, and hence,
(\ref{Heston}) contains a mixed spatial-derivative term.
The Heston PDE is complemented with initial and boundary conditions which are 
determined by the specific option under consideration.
In this paper we shall assume boundary conditions of Dirichlet type.

A widely known semi-discretization of PDEs in finance is given by central 
second-order finite difference (FD) schemes, see e.g.~\cite{TR00,W99}.
To render the numerical solution of the Heston PDE feasible, the spatial domain
is first restricted to a bounded set $[L,S]\times[0,V]$ with fixed values $S$, 
$V$ chosen sufficiently large, with additional Dirichlet conditions imposed at 
$s=S$ and $v=V$.
Let $m_1, m_2 \ge 3$ be any given integers and define spatial mesh widths 
\[
\Delta s = \frac{S-L}{m_1+1}~~,~~\Delta v = \frac{V}{m_2+1}\,.
\]
The central second-order FD schemes for approximating the 
advection, diffusion and mixed derivative terms in (\ref{Heston}) are
\begin{subeqnarray}\label{sv}
\left(u_s\right)_{i,j}
&\approx &  \frac{u_{i+1,j}-u_{i-1,j}}{2\Delta s},\\
\left(u_v\right)_{i,j}
&\approx & \frac{u_{i,j+1}-u_{i,j-1}}{2\Delta v},\\
\left(u_{ss}\right)_{i,j}
&\approx & \frac{u_{i+1,j}-2u_{i,j}+u_{i-1,j}}{(\Delta s)^2},\\
\left(u_{vv}\right)_{i,j}
&\approx & \frac{u_{i,j+1}-2u_{i,j}+u_{i,j-1}}{(\Delta v)^2},\\
\left(u_{sv}\right)_{i,j}
&\approx & \frac{u_{i+1,j+1}+u_{i-1,j-1}-u_{i-1,j+1}-u_{i+1,j-1}}{4\Delta s \Delta v},
\end{subeqnarray}
with the short-hand notation $u_{i,j} = u(s_i, v_j, t)$ and spatial grid
points
\[
s_i=L+i\cdot \Delta s~~(i=0,1,\ldots,m_1+1)~~,~~v_j=j\cdot \Delta v~~(j=0,1,\ldots,m_2+1). 
\]
Semi-discretization by (\ref{sv}) of a given initial-boundary value problem 
for the Heston PDE leads to an initial value problem for a large system of 
ordinary differential equations (ODEs),
\begin{equation}
\label{ODE}
U'(t)=AU(t)+b(t)\quad (0\leq t\leq T),\quad U(0)=U_0.
\end{equation}
Here $A$ is a given constant real $m \times m$ matrix and $b(t)$ (for 
$0\leq t\leq T$) and $U_0$ are given real $m \times 1$ vectors with 
$m= m_1m_2$.
The vector $U_0$ is directly obtained from the initial condition for 
(\ref{Heston}), whereas the vector function $b$ depends on the boundary 
conditions. 
For each $t>0$, the entries of the solution vector $U(t)$ to (\ref{ODE}) form 
approximations to the exact solution values $u(s_i,v_j,t)$ for $1\le i \le m_1$, 
$1\le j \le m_2$.

The aim of our paper is to gain insight into the stability of the semi-discrete 
\mbox{Heston} PDE (\ref{ODE}).
To this purpose, we are interested in the existence of useful, rigorous upper 
bounds on the quantity $||e^{tA}||$ (for $t\ge 0$) where \mbox{$||\,\cdot\,||$} 
denotes an induced matrix norm.
Such bounds, on the magnitude of the matrix exponential of $tA$, guarantee 
that any (rounding or discretization) errors cannot grow excessively.
For central second-order FD discretizations of the Black--Scholes PDE, 
adequate stability bounds were recently proved in \cite{IHV09}.
These bounds are of the well-known type  
\begin{equation}
\label{expA}
||e^{tA}||\leq K e^{t\omega}\quad (t\geq 0)
\end{equation}
with constants $\omega\in\R$ and $K\geq 1$.
To our knowledge stability estimates of the type (\ref{expA}) have not 
been obtained in the literature up to now for FD discretizations of the 
Heston PDE.
In the present paper, we shall establish a natural extension of stability 
results derived in \cite{IHV09}.
We note that a main difficulty in proving this extension lies in the mixed 
derivative term in the Heston PDE, which does not arise in the Black--Scholes 
case.

As the semi-discrete Heston matrix $A$ is in general non-normal, bounds on the norm 
of $e^{tA}$ which are based solely on the eigenvalues of $A$ are most often not useful.
For the stability analysis in this paper, we shall employ the {\it logarithmic 
spectral norm}.
For any given complex $k\times k$ matrix $A$, with integer $k\ge 1$, it is defined 
by the limit
\[
\mu_2[A]=\lim_{t\downarrow0}\frac{||I+t A||_2 -1}{t}\,,
\] 
where $||\,\cdot\,||_2$ is the spectral norm and $I$ is the $k\times k$
identity matrix. 
We note that general complex matrices $A$ are considered for later use.
The following key result forms the basis for our analysis; see e.g.
\cite{HNW08,HV03,So06,TE05}.
\begin{theorem}
\label{stab lognorm}
Let $A$ be any complex $k\times k$ matrix and $\omega\in\R$.
Then
\begin{equation*}
\mu_2 [A]\leq\omega \quad \Longleftrightarrow \quad
||e^{tA}||_2 \leq e^{t\omega}~\textrm{ for all }~ t\geq 0.
\end{equation*}
\end{theorem}
Denote by $\langle \cdot\, , \cdot \rangle_2$ and $|\cdot|_2$ the 
standard inner product and Euclidean norm, respectively.
Then for the logarithmic spectral norm one has the more convenient formulas
\begin{subeqnarray}
\label{mu two}
\mu_2[A] &=&
\max \left\{ \textrm{Re} \langle Ax,x \rangle_2 \,:\, x\in\C^k \,,\, |x|_2 = 1 \right\}\\
&=&
\max\left\{\lambda : \lambda \textrm{ eigenvalue of } \tfrac{1}{2}(A+A^\ast) \right\},
\end{subeqnarray}
where $A^\ast$ stands for the Hermitian adjoint of $A$.

Motivated by the study \cite{IHV09} for the Black--Scholes PDE, we introduce 
also a suitably scaled version of the spectral norm on $\C^{m\times m}$.
Consider the positive diagonal matrices
\[
D_1 = \textrm{diag}(s_1,s_2,\ldots,s_{m_1})\,,~
D_2 = \textrm{diag}(v_1,v_2,\ldots,v_{m_2})\,,~
D = D_2 \otimes D_1\,,
\]
where $\otimes$ is the Kronecker product.
For vectors $x\in \C^m$ we define the norm
\[
|x|_D = |D^{-1/2}\,x|_2
\] 
and denote for matrices $A\in \C^{m\times m}$ the induced matrix norm 
and logarithmic norm by $||A||_D$ and $\mu_D[A]$, respectively.
For any matrix $A$ there holds
\begin{equation}
\label{scaled lognorm}
||A||_D = ||D^{-1/2}\,A\,D^{1/2}||_2~~,~~
\mu_D[A] = \mu_2[D^{-1/2}\,A\,D^{1/2}]
\end{equation}
and the spectral norm of $A$ is bounded in terms of its scaled version through
\begin{equation}
\label{boundDnorm}
||A||_2 \le \sqrt{ \frac{s_{m_1}v_{m_2}}{s_1 v_1} } \cdot ||A||_D \,.
\end{equation}

The outline of the paper is as follows. 
In Section \ref{stabbounds} we derive practical stability bounds for the
semi-discrete Heston PDE (\ref{ODE}).
Here the advection and diffusion terms are each studied individually.
Numerical illustrations are provided in Section~\ref{numexp}, with 
actual computations of the norms of matrix exponentials.
Conclusions and issues for future research are discussed in 
Section~\ref{concl}.

\setcounter{equation}{0}
\setcounter{theorem}{0}
\section{Stability bounds}
\label{stabbounds}
Let $I$ denote the identity matrix of generic dimension.
Associated with the~FD formulas (\ref{sv}), we define the tridiagonal 
$m_1\times m_1$ matrices
\[
L_1=\frac{1}{2\Delta s}\cdot\textrm{tridiag}\left(-1\,,\,0\,,\,1\right)~~,~~
M_1=\frac{1}{(\Delta s)^2}\cdot\textrm{tridiag}\left(1\,,\,-2\,,\,1\right)\phantom{.}
\]
and the tridiagonal $m_2\times m_2$ matrices
\[
L_2=\frac{1}{2\Delta v}\cdot\textrm{tridiag}\left(-1\,,\,0\,,\,1\right)~~,~~
M_2=\frac{1}{(\Delta v)^2}\cdot\textrm{tridiag}\left(1\,,\,-2\,,\,1\right).
\]
FD discretization by (\ref{sv}) of the spatial derivative terms $rsu_s$, 
$\kappa(\eta-v)u_v$, $\tfrac{1}{2}s^2vu_{ss}$, $\rho\sigma svu_{sv}$, 
$\tfrac{1}{2}\sigma^2vu_{vv}$ in the Heston PDE (\ref{Heston}) gives
rise to the following real $m\times m$ matrices, respectively:
\begin{subeqnarray}\label{Ak}
A_1&=&rI\otimes(D_1L_1)\,,\\
A_2&=&\kappa[(\eta I-D_2)L_2]\otimes I\,,\\
A_3&=&\tfrac{1}{2}D_2\otimes(D_1^2M_1)\,,\\
A_4&=&\rho\sigma(D_2L_2)\otimes(D_1L_1)\,,\\
A_5&=&\tfrac{1}{2}\sigma^2(D_2M_2)\otimes I\,.
\end{subeqnarray}
Here a lexicographic ordering of the spatial grid points is considered.
It is worth noting that $(D_2L_2)\otimes(D_1L_1)$ in (\ref{Ak}d) can 
be regarded as a discrete analogue of $(vu_v) \circ (su_s) = svu_{sv}$ 
where $\circ$ denotes composition.
The semi-discrete Heston matrix $A$ in (\ref{ODE}) is equal to
\[
A = A_1+A_2+A_3+A_4+A_5-rI.
\]

Our introductory result concerns the two parts of the semi-discrete Heston 
matrix corresponding to the advection terms in the $s$- and $v$-directions. 
It provides useful stability bounds of the type (\ref{expA}) for these.
\begin{theorem}
Let $r, \kappa, \eta >0$ and let $A_1$, $A_2$ be given by (\ref{Ak}a), 
(\ref{Ak}b). Then 
\[
||e^{tA_1}||_2\leq 
e^{t\omega}\quad (t\geq 0) ~~~~{\it with}~~\omega=\frac{r}{2}\phantom{.}
\]
and 
\[
||e^{tA_2}||_2\leq 
e^{t\omega}\quad (t\geq 0) ~~~~{\it with}~~\omega=\frac{\kappa}{2}.
\]
The above values for $\omega$ are the smallest that hold uniformly in the 
respective mesh widths.
\end{theorem}
\noindent\textbf{Proof}
Consider the symmetric matrix 
$F=\textrm{tridiag}\left(\tfrac{1}{2},0,\tfrac{1}{2}\right)$ 
of generic dimension.
All eigenvalues of this matrix lie in the real interval $[-1,1]$. 
It is readily verified that $A_1+A_1^\rT = -r I \otimes F$
and $A_2+A_2^\rT = \kappa F \otimes I$ (for any $\eta$).
The eigenvalues of $I \otimes F$ and $F \otimes I$ are the same 
as those of the pertinent matrix $F$. 
By (\ref{mu two}b) it thus follows that $\mu_2[A_1] \le r/2$ and 
$\mu_2[A_2] \le \kappa/2$ and application of Theorem \ref{stab lognorm} 
yields the required bounds.
Furthermore, as there exist eigenvalues of $F$ that converge to $-1$ and 
$1$ when the dimension increases, the obtained values for $\omega$ are 
the smallest that hold uniformly in the respective mesh widths. 
\begin{flushright}
$\Box$
\end{flushright}

The subsequent lemma deals with the logarithmic spectral norm 
of certain matrices of block Toeplitz type and is essential to 
the proof of our main result in this paper.
Let $E=\textrm{tridiag}(0,0,1)$ denote the $m_2\times m_2$ 
forward shift matrix.

\begin{lemma}\label{toeplitz} 
Let $B_0$, $B_1$ be any given real $m_1\times m_1$ matrices and let 
the $m\times m$ matrix $B$ be defined by
\[
B = I\otimes B_0 + E\otimes B_1 +E^\rT \otimes B_1^\rT.
\] 
Then 
\[
\mu_2[B] \leq \max_{\zeta\in\C,|\zeta|=1} \mu_2[B_0+2\zeta B_1].
\]
\end{lemma}
\noindent\textbf{Proof}
Consider the so-called symbol of $B$, given by
\[
B(\zeta) = B_0 + \zeta B_1 + \zeta^{-1}B_1^\rT 
\]
for $\zeta\in\C$, $|\zeta|=1$.
Since $B$ is a block Toeplitz matrix, also the exponential
\[
e^{tB} = \sum_{j=0}^\infty\, \frac{t^j}{j!}\, B^j
\]
is block Toeplitz.
The symbol of $e^{tB}$ is equal to $e^{tB(\zeta)}$ and one has the 
bound
\[
||e^{tB}||_2 \le \max_{\zeta\in\C,|\zeta|=1} ||e^{tB(\zeta)}||_2\,,
\]
which is a consequence of Parseval's identity, see e.g.~\cite[p.186]{BS99}.
By Theorem \ref{stab lognorm} it readily follows from this that
\[
\mu_2[B] \leq \max_{\zeta\in\C,|\zeta|=1} \mu_2[B(\zeta)].
\]
Let $\widehat{B}(\zeta) = B_0 + 2 \zeta B_1$. 
Then the Hermitian parts of $B(\zeta)$ and $\widehat{B}(\zeta)$ are equal
and hence, by (\ref{mu two}b), there holds 
$\mu_2[B(\zeta)] = \mu_2[\widehat{B}(\zeta)]$.
This yields the proof.
\begin{flushright}
$\Box$
\end{flushright}

Our main result of this paper concerns the stability of the diffusion part
(including the mixed derivative term) of the semi-discrete Heston system.
\begin{theorem}
\label{stabdiff}
Let $\sigma>0$ and $\rho\in[-1,1]$ and let $A_3$, $A_4$, $A_5$ 
be given by (\ref{Ak}c), (\ref{Ak}d), (\ref{Ak}e). 
Then, for all $t\ge 0$,
\begin{subeqnarray}
\label{expdiff}
||e^{t(A_3+A_4+A_5)}||_D&\leq&1\,,\\
||e^{t(A_3+A_4+A_5)}||_2\,&\leq&\sqrt{ \frac{s_{m_1}v_{m_2}}{s_1 v_1} }\,. 
\end{subeqnarray}
\end{theorem}

\noindent
The strong stability result (\ref{expdiff}a) means that the diffusion 
part of the semi-discrete Heston system is contractive in the scaled 
spectral norm.
The bound (\ref{expdiff}b) for the standard spectral norm 
is discussed in more detail in Section~\ref{numexp}.
Theorem \ref{stabdiff} can be viewed as a natural extension of 
\cite[Theorem 2.8]{IHV09} that was derived for the case of the 
Black--Scholes PDE.
In the special situation where $\rho =0$, so that no mixed derivative 
term is present in the Heston PDE and the matrix $A_4$ vanishes, the 
result of Theorem \ref{stabdiff} can be obtained in analogous way to 
loc.~cit.
However, the important general situation where $\rho \not=0$ requires
a new, and more elaborate, proof.

\bigskip\noindent\textbf{Proof}
The bound (\ref{expdiff}b) follows directly from (\ref{expdiff}a) by 
(\ref{boundDnorm}).
By Theorem \ref{stab lognorm}, the bound (\ref{expdiff}a) is 
equivalent to $\mu_{D}[A_3+A_4+A_5]\leq 0$. 
In the following we show that this condition holds.
For convenience, the proof is split into three, consecutive parts.
\\
\\
{(i)}~For any given real square matrices $A$, $G$ with $G$ nonsingular 
it holds that $\mu_2[A]\leq 0$ if and only if $\mu_2[G^{\rm T} A\, G]\leq 0$.
Choosing $A=A_3+A_4+A_5$ and $G=D_2^{-1/2}\otimes I$, and taking into account 
(\ref{scaled lognorm}), we obtain
\[
\mu_{D}[A_3+A_4+A_5] \leq 0 ~~~\Longleftrightarrow~~~ 
\mu_2[B] \leq 0,
\]
where the matrix $B$ is given by
\begin{eqnarray*}
B 
&=& 
(D_2^{-1}\otimes D_1^{-1/2})(A_3+A_4+A_5)(I\otimes D_1^{1/2})\\
&=& 
\tfrac{1}{2}I\otimes(D_1^{3/2} M_1 D_1^{1/2})
+\rho\sigma L_2\otimes(D_1^{1/2} L_1 D_1^{1/2})
+\tfrac{1}{2}\sigma^2 M_2 \otimes I. 
\end{eqnarray*}
Let $\ts = \sigma/\Delta v$ and define the matrices
\[
\tL_1 = D_1^{1/2} L_1 D_1^{1/2}~~,~~\tM_1 = D_1^{3/2} M_1 D_1^{1/2}.
\]
Note that
\[
L_2=\frac{1}{2\Delta v}(E-E^\rT)~~,~~M_2=\frac{1}{(\Delta v)^2}(E-2I+E^\rT).
\]
Inserting into $B$ yields
\begin{eqnarray*}
B
&=&\tfrac{1}{2} I \otimes \tM_1 + \rho \sigma L_2 \otimes 
\tL_1 + \tfrac{1}{2} \sigma^2 M_2 \otimes I\\
&=&\tfrac{1}{2} \left[ I \otimes \tM_1 + \rho \ts (E-E^\rT ) \otimes 
\tL_1 +b\ts^2 (E-2I+E^\rT )\otimes I \right]\\
&=&\tfrac{1}{2} \left[ I \otimes (\tM_1 -2\ts^2I) +E \otimes 
(\rho \ts \tL_1 +\ts^2 I) +E^\rT \otimes 
(-\rho \ts \tL_1 +\ts^2 I) \right].
\end{eqnarray*}
Since $\tL_1^\rT =-\tL_1$ we are in the situation of Lemma 
\ref{toeplitz}. 
Application of this lemma yields the following sufficient condition for 
$\mu_2[B]\leq 0$, with $\zeta\in\C\,$:
\begin{equation*}
\mu_2\left[\tfrac{1}{2}\tM_1-\ts^2I+\zeta(\rho\ts\tL_1+\ts^2I)\right]\leq 0
~~\textrm{whenever}~|\zeta|=1.
\end{equation*}
Let $\imi$ denote the imaginary unit and let $\lm [A]$ stand for
the maximum eigenvalue of any matrix $A$ having just real eigenvalues. 
Using (\ref{mu two}b) one readily finds that the sufficient condition 
above is equivalent to
\begin{equation}\label{suff1}
\lm\left[\tfrac{1}{2}(\tM_1+\tM_1^\rT )+
2\imi(\textrm{Im\,}\zeta)\rho\ts \tL_1\right]
\leq 2\ts^2(1-\textrm{Re\,}\zeta)
~~\textrm{whenever}~|\zeta|=1.~
\end{equation}
\\
{(ii)}~Define $C_s = D_1 L_1$ and $C_{ss} = \tfrac{1}{2}D_1^2 M_1$.
Remark that these matrices can be viewed as FD discretizations of the 
$su_s$ and $\tfrac{1}{2} s^2 u_{ss}$ terms, respectively.
Clearly,
\[
\tL_1 = D_1^{-1/2} C_s D_1^{1/2}.
\]
Next, a direct calculation shows that
\[
\tfrac{1}{2} (M_1 D_1 - D_1 M_1) = L_1
\]
and using this one obtains
\[
\tfrac{1}{2}(\tM_1+\tM_1^\rT )= D_1^{-1/2} (2C_{ss}+C_s) D_1^{1/2}.
\]
Therefore, by a similarity transformation, (\ref{suff1}) is equivalent 
to
\begin{equation}\label{suff2}
\lm\left[ C_{ss}+\tfrac{1}{2}C_s + 
\imi (\textrm{Im\,}\zeta)\rho\ts C_s \right] \leq 
\ts^2(1-\textrm{Re\,}\zeta) ~~\textrm{whenever}~|\zeta|=1.
\end{equation}
For $|\zeta|=1$ it holds that
\[
1-\textrm{Re\,}\zeta \ge 
\tfrac{1}{2}(1+\textrm{Re\,}\zeta)(1-\textrm{Re\,}\zeta) =
\tfrac{1}{2}(1-(\textrm{Re\,}\zeta)^2) =
\tfrac{1}{2}(\textrm{Im\,}\zeta)^2.
\]
This bound gives rise to the following sufficient 
condition:
\[
\lm\left[ C_{ss}+\tfrac{1}{2}C_s+\imi y\rho\ts C_s \right]
\leq \tfrac{1}{2}\ts^2y^2~~\textrm{whenever}~y\in\R, |y|\leq 1.
\]
Then, upon replacing $\tfrac{1}{2}y\rho\ts$ by $y$ and using 
that the correlation $\rho$ satisfies $|\rho|\leq 1$, we arrive 
at the neat condition
\begin{equation}\label{suff3}
\lm\left[C_{ss}+\left(\tfrac{1}{2}+2\imi y\right)C_s\right]
\leq 2y^2~~\textrm{whenever}~y\in\R.
\end{equation}
Summarizing,
\[
(\ref{suff3}) ~~\Longrightarrow~~ (\ref{suff2}) ~~\Longleftrightarrow~~ 
(\ref{suff1}) ~~\Longrightarrow~~ \mu_2[B] \leq 0.
\]
In the third and final part we prove that (\ref{suff3}) is fulfilled.
\\
\\
{(iii)}~Let $\mu_\infty [A]$ denote the logarithmic maximum norm of 
any complex square matrix $A$.
It is well-known that if $A=(a_{i,j})$ then
\begin{equation*}
\label{mu inf}
\mu_\infty[A] = 
\max_i\,(\, \textrm{Re}\, a_{i,i}+\sum_{j\neq i}|a_{i,j}| \,)\,.
\end{equation*}
Any induced logarithmic norm forms an upper bound on the real parts 
of the eigenvalues of $A$.
In the following, the logarithmic maximum norm will be used to 
this purpose.

Write $\nu_i = s_i/\Delta s$ for $1\le i \le m_1$.
There holds
\[C_{ss}+\left(\tfrac{1}{2}+2\imi y\right)C_s
=\textrm{tridiag}(\beta_i,\alpha_i,\gamma_i)\] with
\[
\alpha_i = -\nu_i^2~~,~~
\beta_i  = \tfrac{1}{2}\nu_i\left(\nu_i-\tfrac{1}{2}-2\imi y\right)~~,~~
\gamma_i = \tfrac{1}{2}\nu_i\left(\nu_i+\tfrac{1}{2}+2\imi y\right).
\]
In proving (\ref{suff3}) we need to distinguish two cases: $|y|\geq 1/2$, 
and the more intricate case $|y|<1/2$.
\\
\\
\framebox{$|y|\geq 1/2$}~~~Put $\beta_1=0$, $\gamma_{m_1}=0$. One has
\[
\lm\left[C_{ss}+\left(\tfrac{1}{2} +2\imi y\right)C_s\right]
\leq\mu_{\infty}\left[C_{ss}+\left(\tfrac{1}{2} +2\imi y\right)C_s\right]=
\max_{1\leq i\leq m_1} \{\alpha_i+|\beta_i|+|\gamma_i|\}.
\]
Let $1\leq i\leq m_1$.
Then
$\alpha_i+|\beta_i|+|\gamma_i|\leq 2y^2$ if 
\[
\nu_i \sqrt{\left(\nu_i-\tfrac{1}{2}\right)^2+\theta\,} \,+\,
\nu_i \sqrt{\left(\nu_i+\tfrac{1}{2}\right)^2+\theta\,}\leq 2\nu_i^2+\theta,
\]
where $\theta=4y^2$. 
By an elementary calculation one verifies that this inequality is 
equivalent to
\[
4\theta(\theta-1)\nu_i^4+\theta^2(4\theta-1)\nu_i^2+\theta^4\geq 0,
\]
which holds whenever $\theta\geq 1$.
Thus, condition (\ref{suff3}) is valid whenever $|y|\geq 1/2$.
\\
\\
\framebox{$|y|< 1/2$}~~~Let 
$\Delta=\textrm{diag}(\delta_1,\delta_2,\ldots,\delta_{m_1})$ with
arbitrary real numbers $\delta_i > 0$ ($1\leq i\leq m_1$) and write 
$\varepsilon_i=\delta_i/\delta_{i-1}$ ($2\leq i\leq m_1$). 
A similarity transformation with the diagonal matrix $\Delta$ leads 
to the following bound,
\begin{align*}
&\lm\left[C_{ss}+\left(\tfrac{1}{2}
+2\imi y\right)C_s\right]\leq
\mu_{\infty}\left[\Delta\left(C_{ss}
+\left(\tfrac{1}{2}+
2\imi y\right)C_s\right)\Delta^{-1}\right]=\\
&\max\left\{\alpha_1+\frac{1}{\varepsilon_2}|\gamma_1|\,,\,
\max_{2\leq i\leq m_1-1}\{\alpha_i+\varepsilon_i|\beta_i|
+\frac{1}{\varepsilon_{i+1}}|\gamma_i|\}\,,\,
\alpha_{m_1}+\varepsilon_{m_1}|\beta_{m_1}|\right\}.
\end{align*}
Let $2\leq i\leq m_1-1$. 
The estimate 
\[
|x \pm 2 \imi y| \leq x+\frac{2y^2}{x} 
~~~(\textrm{for}~x \in\R, x>0)
\]
yields
\[
\alpha_i+\varepsilon_i|\beta_i|+\frac{1}{\varepsilon_{i+1}}|\gamma_i|
\leq a_i+b_i\cdot2y^2,
\]
where
\begin{eqnarray*}
a_i&=&
\frac{\nu_i}{2} \left[-2\nu_i +\varepsilon_i (\nu_i -\tfrac{1}{2})
+\frac{1}{\varepsilon_{i+1}} (\nu_i +\tfrac{1}{2}) \right],\\
b_i&=&
\frac{\nu_i}{2} \left[\frac{\varepsilon_i}{\nu_i -\tfrac{1}{2}}
+\frac{1}{\varepsilon_{i+1}(\nu_i +\tfrac{1}{2})}\right].
\end{eqnarray*}
It holds that
\[
a_i+b_i\cdot2y^2\leq 2y^2~~\left(\textrm{whenever 
$|y|<\tfrac{1}{2}$}\right)~~\Longleftrightarrow~~ 
a_i\leq 0 ~~\textrm{and}~~ 2a_i+b_i\leq 1.
\]
By trial and error, we have found the convenient choice
\begin{equation}\label{epsil}
\varepsilon_j=\frac{(\nu_j -\tfrac{1}{2})(\nu_j +\tfrac{1}{2})}{\nu_j^2}
~~~(\textrm{for}~2\leq j\leq m_1).
\end{equation}
Using (\ref{epsil}), and noticing that $\nu_{i+1} = \nu_i+1$, it is easily 
seen that
\[
a_ i = -\frac{1}{8} \frac{\nu_i-\tfrac{3}{4}}{\nu_i (\nu_i+\tfrac{3}{2})}\,.
\]
Since $\nu_i\ge i\ge 2$ there follows $a_i < 0$. 
Next,
\[
b_i = \frac{\nu_i}{2} \left[ \frac{\nu_i+\tfrac{1}{2}}{\nu_i^2} +
\frac{(\nu_i+1)^2}{(\nu_i+\tfrac{1}{2})^2 (\nu_i+\tfrac{3}{2})} \right].
\]
A straightforward calculation shows that
\[
2a_i+b_i\leq 1 ~~\Longleftrightarrow~~ 
\nu_i^3 - \tfrac{3}{4}\nu_i^2 - \tfrac{3}{2}\nu_i - \tfrac{9}{16} \ge 0.
\]
It is readily verified that the inequality in the right-hand side is 
fulfilled.
Hence, with (\ref{epsil}),
\[
\alpha_i+\varepsilon_i|\beta_i|+\frac{1}{\varepsilon_{i+1}}|\gamma_i| \le
2y^2~~~(2\leq i\leq m_1-1).
\]
By an analogous reasoning it follows that
\[
\alpha_1+\frac{1}{\varepsilon_2}|\gamma_1|  \leq 2y^2~~,~~
\alpha_{m_1}+\varepsilon_{m_1}|\beta_{m_1}| \leq 2y^2.
\]
Consequently, condition (\ref{suff3}) is also valid whenever 
$|y|< 1/2$. This completes the proof of the theorem.
\begin{flushright}
$\Box$
\end{flushright}

\section{Numerical experiments}
\label{numexp}
In this section we numerically examine the stability bound (\ref{expdiff}b) 
of Theorem \ref{stabdiff} for the diffusion part of the semi-discrete Heston 
system.
Slightly rewritten, it reads
\[
||e^{t(A_3+A_4+A_5)}||_2 \leq \sqrt{\frac{L+m_1S}{m_1L+S}\, m_2}~.
\]
The right-hand side is equal to $\sqrt{m_1m_2}=\sqrt{m}$\, if $L=0$, and it 
is at most equal to $\sqrt{\min\{m_1, S/L\}\cdot m_2}$\, whenever $L>0$.
For $L>0$ the stability bound (\ref{expdiff}b) is thus more favorable than 
for $L=0$. 

\begin{figure}
\begin{center}
\begin{tabular}{cc}
\includegraphics[width=0.5\textwidth]{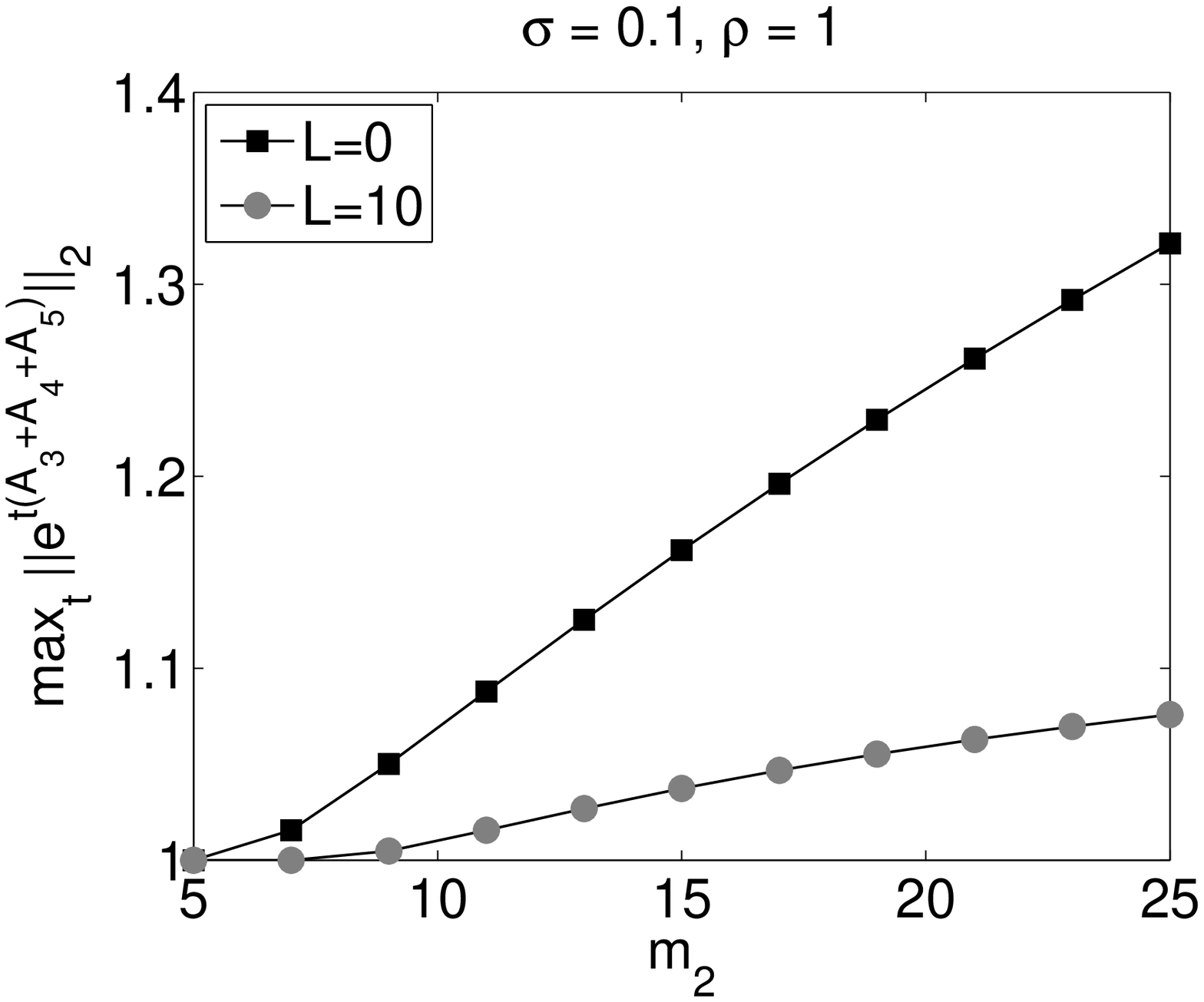}
&\includegraphics[width=0.5\textwidth]{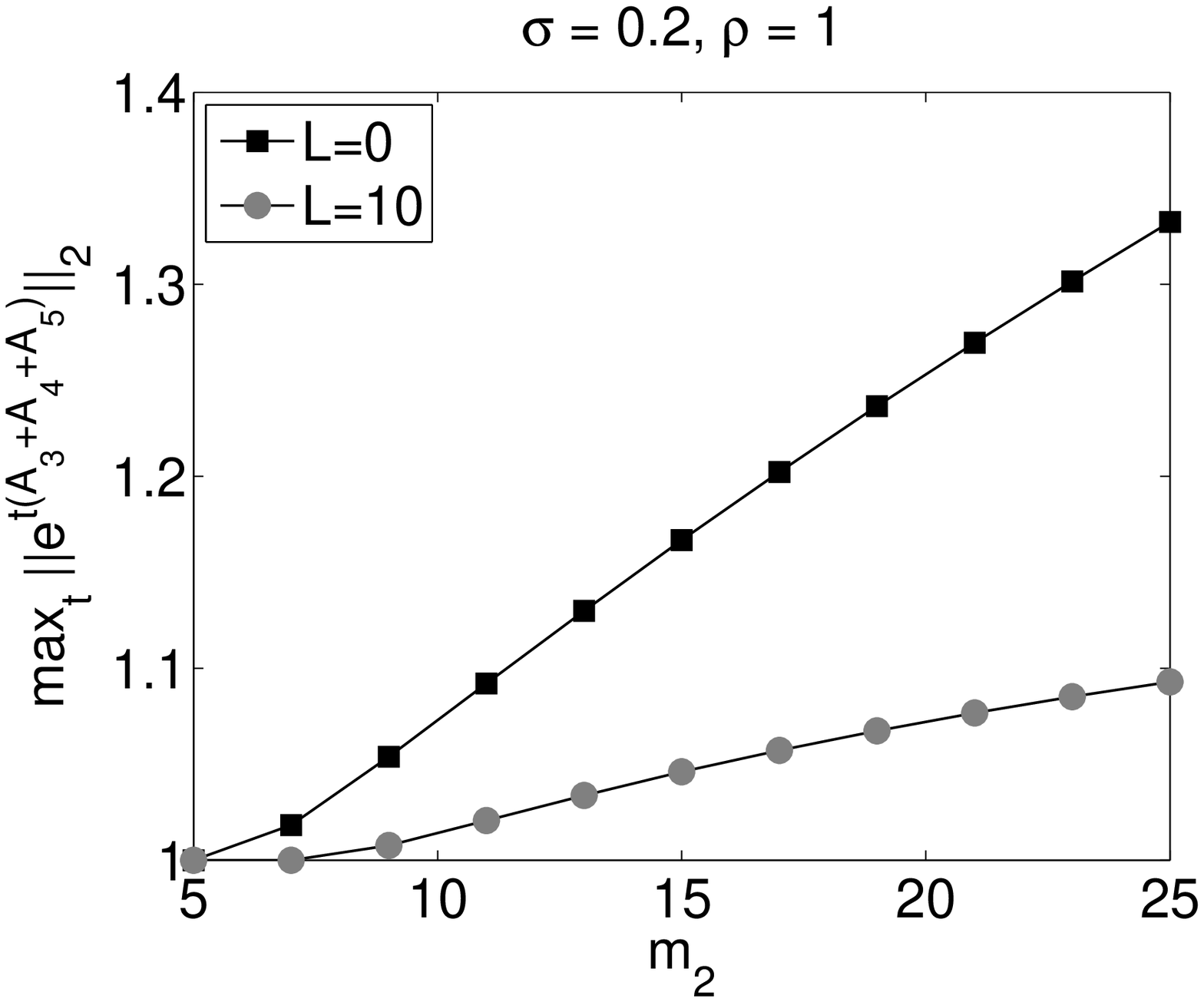}\\
\includegraphics[width=0.5\textwidth]{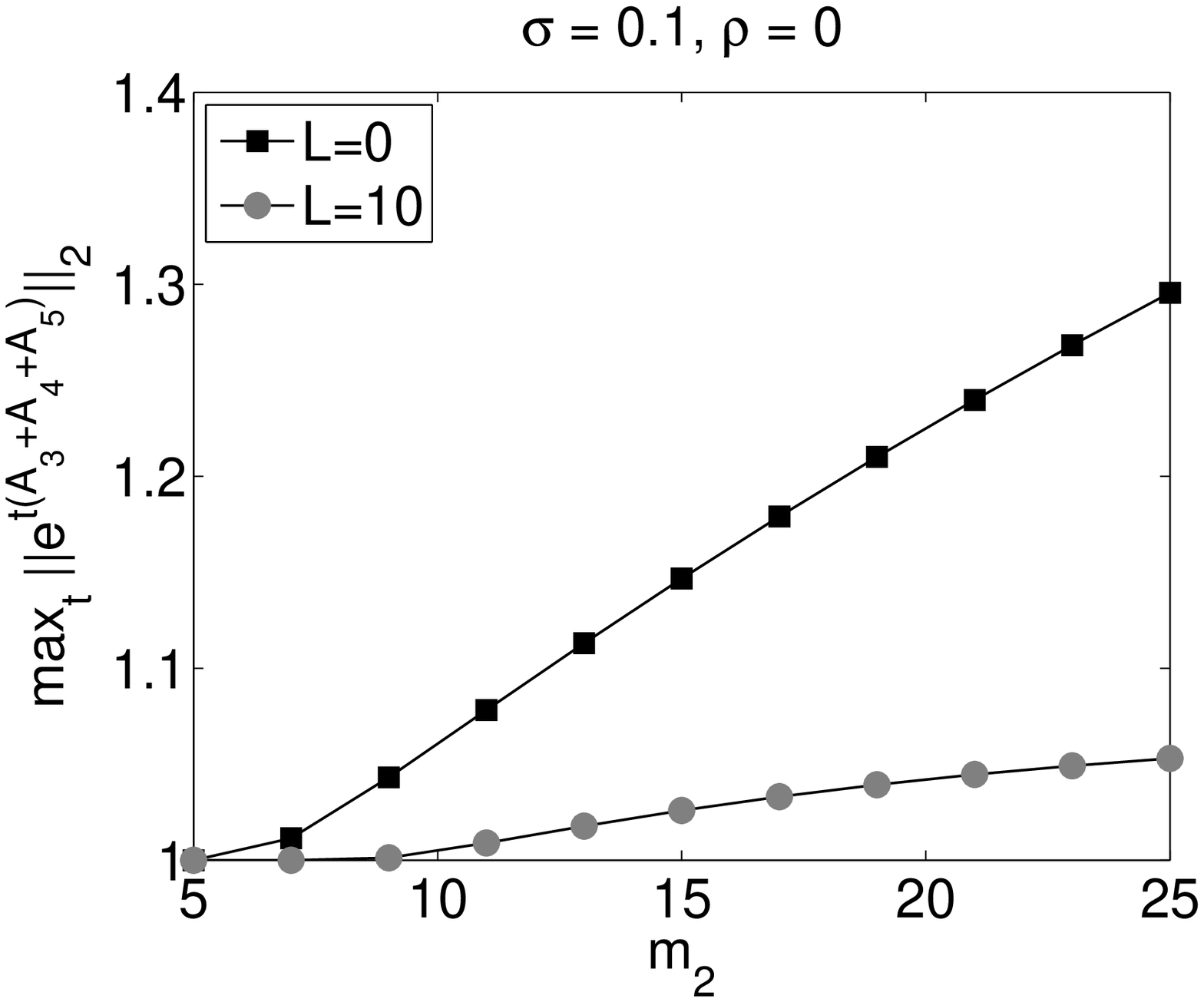}
&\includegraphics[width=0.5\textwidth]{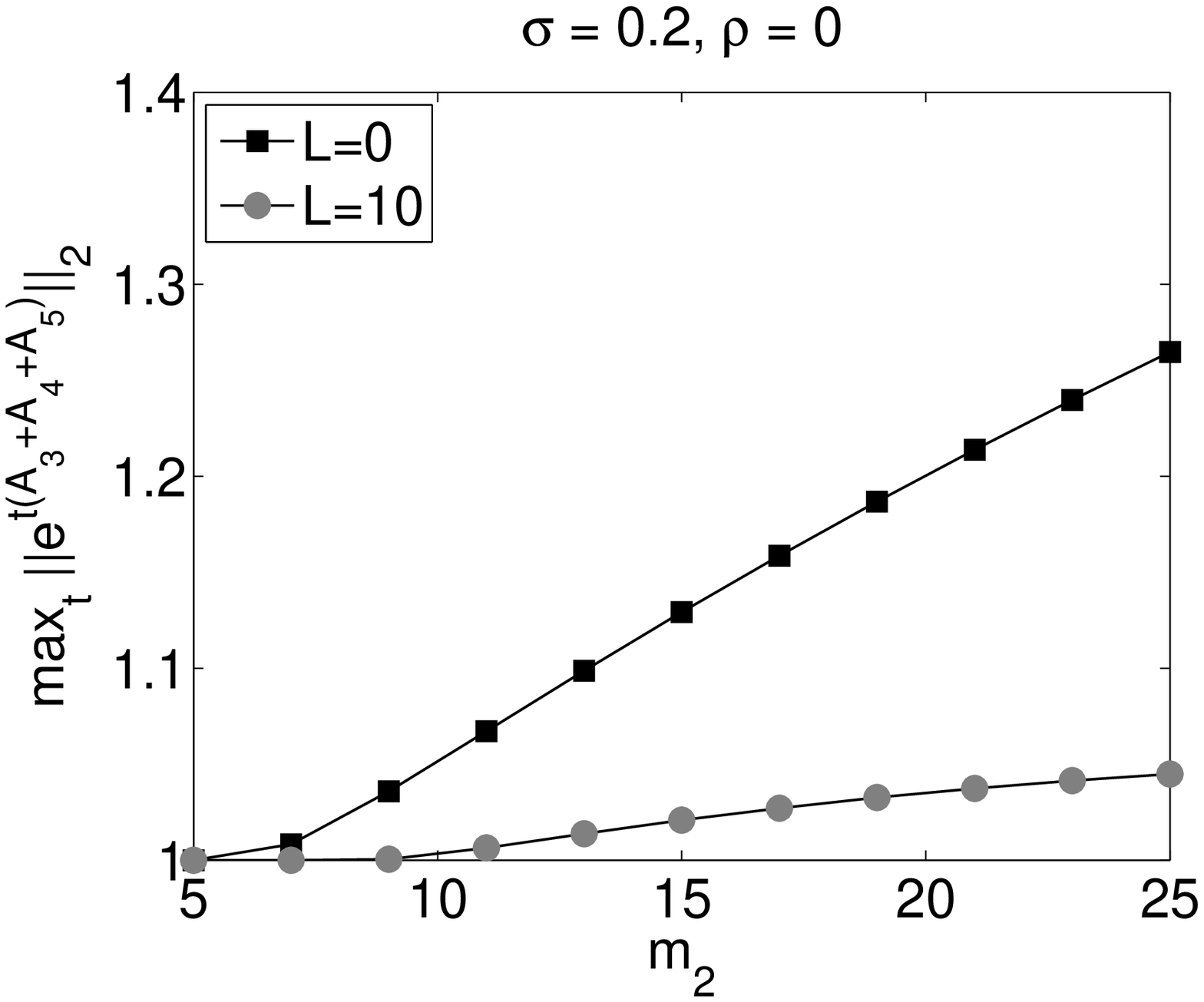}\\
\includegraphics[width=0.5\textwidth]{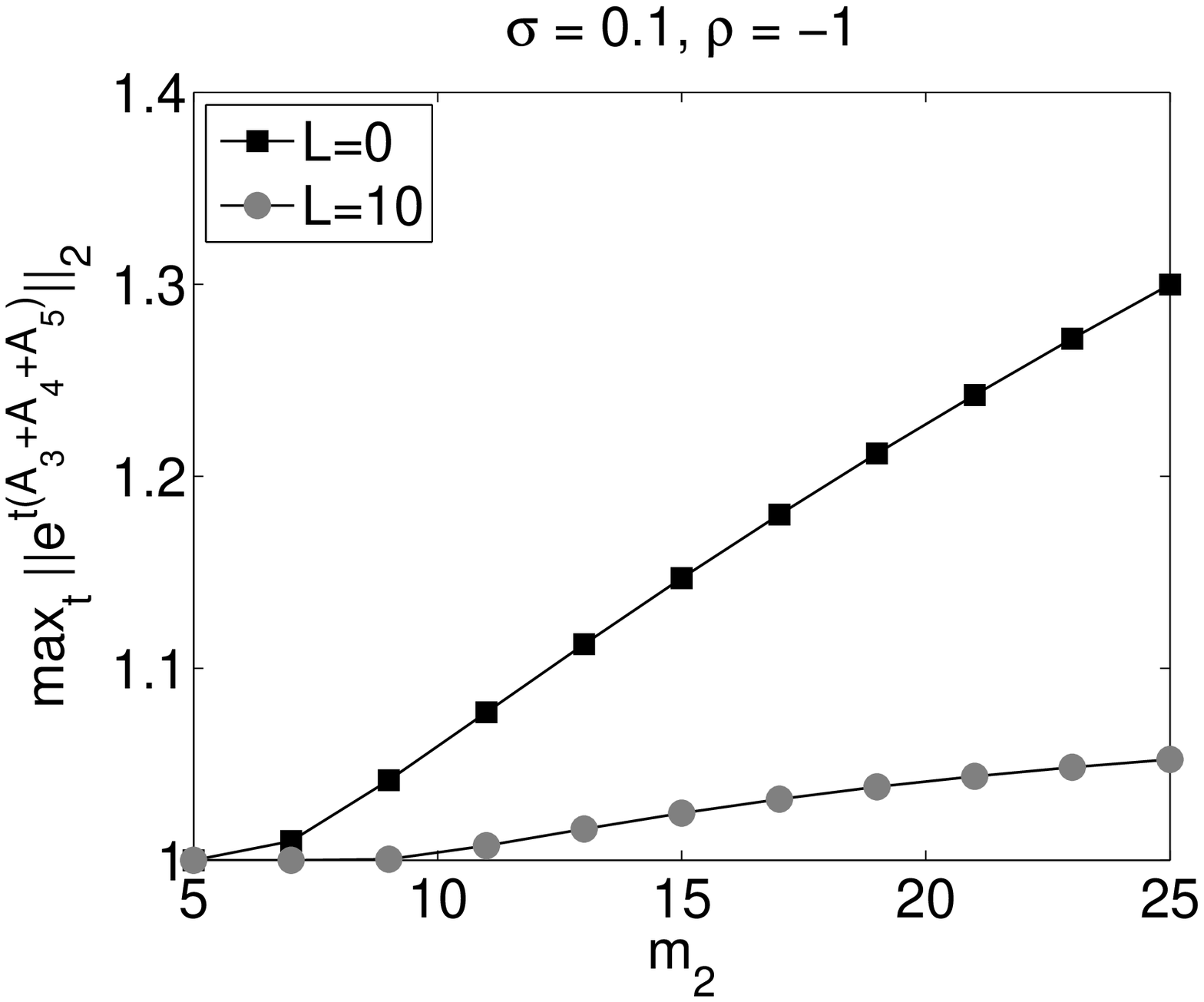}
&\includegraphics[width=0.5\textwidth]{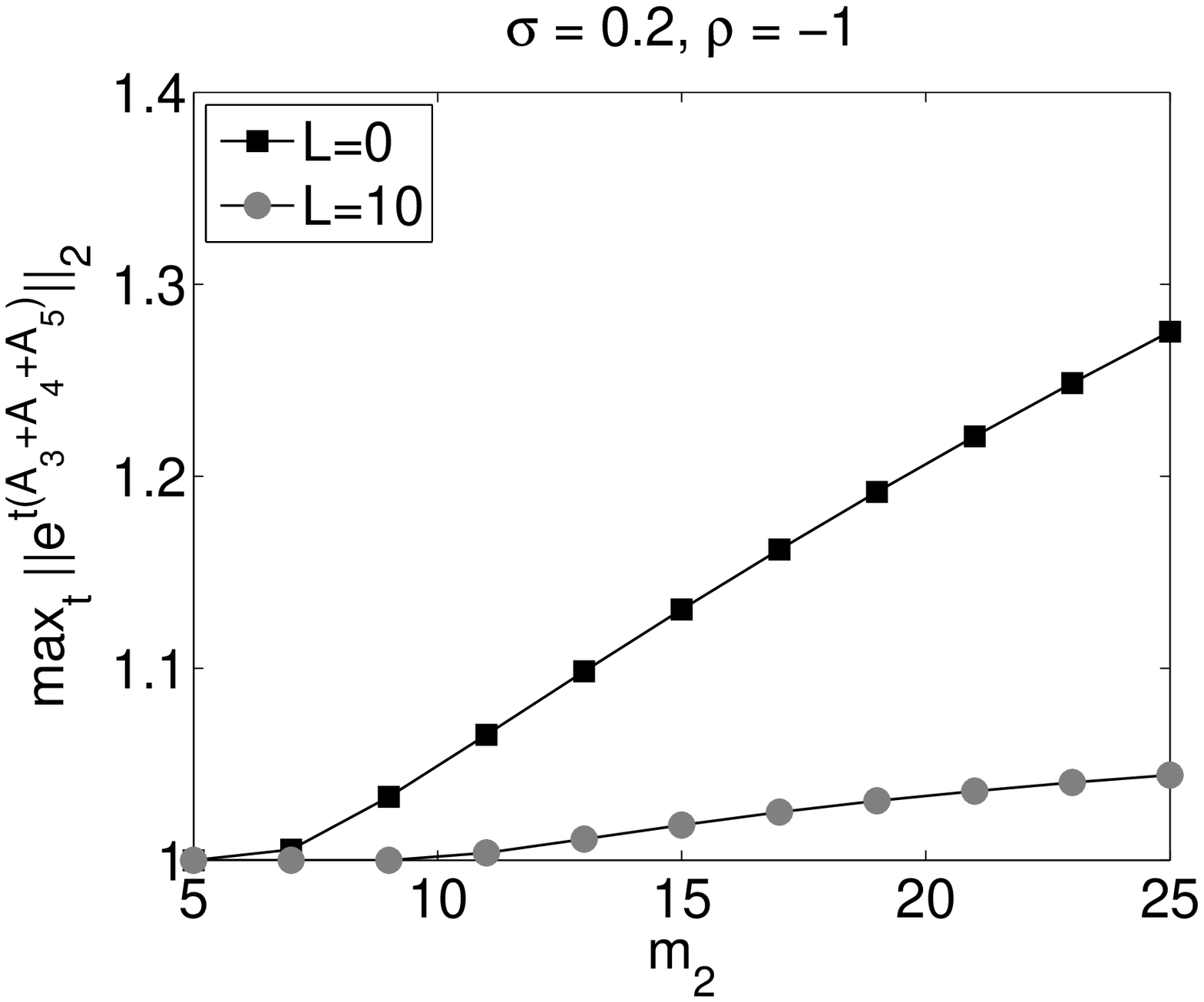}
\end{tabular}
\caption{Graph of estimated $\max_{t\ge 0}\, ||e^{t(A_3+A_4+A_5)}||_2$
vs. $m_2=5,7,9,\ldots,25$ for $L=0$ (black squares) and $L=10$ (grey
circles) where $m_1=2m_2$. 
Left column: $\sigma=0.1$. Right column: $\sigma=0.2$. 
Top row: $\rho=1$. Middle row: $\rho=0$. Bottom row: $\rho=-1$.}
\label{fig1}
\end{center}
\end{figure}

We estimated in MATLAB (version R2009a) the maximum of $||e^{t(A_3+A_4+A_5)}||_2$ 
over $t\geq 0$ for a variety of cases.
We considered all combinations of parameter values 
\[
\sigma \in \{0.1, 0.2\}~~,~~\rho \in \{-1, 0, 1\}~~,~~L \in \{0, 10\}.
\]
Following \cite{IHF10} we chose $m_1=2m_2$ (so that the dimension $m=2m_2^2$)
and selected $m_2=5,7,9,\ldots,25$.
Further $S=800$, $V=5$ were taken as in loc.~cit.
For the computation of the matrix exponential and the spectral norm the 
MATLAB functions {\tt expm} and {\tt norm($\cdot$,2)} were used.
We note that the feasibility of {\tt expm} implied $m_2=25$ as the largest 
reasonable choice (then $m=1250$).
The maximum over $t\ge 0$ was estimated in a basic way by sampling the 
values for $t=0,1,2,\ldots,100$ and subsequently refining in the region 
around the largest value.
We mention that the location of the maximum was always found to lie in 
the interval $0\le t\le 5$.

The obtained results are displayed in Fig.~\ref{fig1}.
Each of the six subfigures shows the estimated maximum of $||e^{t(A_3+A_4+A_5)}||_2$ 
over $t\geq 0$ versus $m_2$ for a given pair $(\sigma, \rho)$.
The black squares correspond to $L=0$ and the grey circles to $L=10$.
As a first observation from Fig.~\ref{fig1}, it is readily seen that all numerical 
results are in agreement with the theoretical stability bound (\ref{expdiff}b).
Secondly, Fig.~\ref{fig1} reveals that for \mbox{$L=10$} the computed maximum 
of $||e^{t(A_3+A_4+A_5)}||_2$ is never larger, and in general much smaller, than 
that for $L=0$.
In addition, we find in all cases a growth that appears to be at most directly 
proportional to $\sqrt{m}\sim m_2$\, and $\sqrt{m_2}$\, when $L=0$ and $L=10$, 
respectively.
This agrees with the bound (\ref{expdiff}b) as well, as discussed above.
Thirdly, Fig.~\ref{fig1} indicates the positive result that the value of 
$\sigma$ and especially $\rho$ only has a limited impact on the actual 
maximum of $||e^{t(A_3+A_4+A_5)}||_2$. 
Note that for $\rho$ we considered here the interesting extreme cases $-1$, 
$0$, $1$, but this result was confirmed by numerical experiments with various 
other values.

\section{Conclusions and future research}
\label{concl}
In this paper useful, rigorous stability bounds have been derived relevant 
to central second-order finite difference discretizations of the Heston PDE 
from mathematical finance.
Results for the advection and diffusion parts have been proved individually
and are valid for arbitrary Heston parameters.
The stability estimates obtained in this paper can be viewed as natural 
extensions of recent stability results from \cite{IHV09} for the case of 
the one-dimensional Black--Scholes PDE.

Besides the standard spectral norm, a suitably scaled version has been considered, 
following a fruitful idea from loc.~cit.
The main result of our paper, Theorem \ref{stabdiff}, states that in this scaled 
spectral norm the semi-discrete diffusion part of the Heston PDE is contractive.
This result holds for arbitrary correlation values $\rho \in [-1,1]$ and thus 
covers the practically important situation where a mixed spatial-derivative term 
is present.

The bound in the standard spectral norm is (also) uniform in $\rho$, which has
been illustrated by ample numerical experiments.
Both theoretical and numerical evidence reveals that in the standard spectral 
norm the stability of the semi-discrete diffusion part is much more favorable 
if the lower barrier $L>0$ than if $L=0$.
In actual applications, $L>0$ is often fulfilled, for example for barrier options; 
else it is harmless to increase $L$ slightly, when the actual region of interest 
for the asset prices lies far away from this value.

We note that the results in this paper can directly be combined, using a well-known 
theorem due to von Neumann \cite[Sects.~IV.11, V.7]{HW02}, to arrive at stability 
bounds for various classes of time-discretization schemes applied to the semi-discrete 
Heston PDE, e.g.~Runge--Kutta methods and linear multistep methods.
For the sake of brevity we have not explicitly included these results here.

In future research we shall investigate, among others, the stability of FD 
schemes for the Heston PDE on non-uniform spatial grids.
Such grids play an important role in mathematical finance.
In \cite{IHV09,V10} stability bounds pertinent to non-uniform grids were derived 
for the case of the Black--Scholes PDE and more general one-dimensional 
advection-diffusion-reaction equations.
In future research we also intend to study for example the adaptation of the 
obtained stability results to different types of boundary conditions.

\section*{Acknowledgments}
The second author acknowledges financial support by the Research Foundation -- 
Flanders, FWO contract no. 1.1.161.10.N.

\end{document}